\documentclass[usenatbib,a4paper,times,fleqn]{mnras}
\usepackage[T1]{fontenc}
\usepackage{graphicx,natbib,times,amssymb,amsmath,pstricks,mathtools,enumitem,array}
\usepackage{aas_macros}
\usepackage{bm,url}
\usepackage{textcomp}
\usepackage{breakurl}
\usepackage{subfig}
\usepackage{nicefrac,xfrac}
\usepackage{mathtools}
\usepackage{cancel}
\usepackage{listings}

\usepackage{cuted}

\newcommand {\apgt} {\ {\raise-.5ex\hbox{$\buildrel>\over\sim$}}\ }
\newcommand {\aplt} {\ {\raise-.5ex\hbox{$\buildrel<\over\sim$}}\ }

\title[CFL condition for the coagulation equation]{On the Courant-Friedrichs-Lewy condition for numerical solvers of the coagulation equation}

\author[Laibe \& Lombart]{Guillaume Laibe$^{1,2}$\thanks{guillaume.laibe@ens-lyon.fr}, Maxime Lombart$^{1}$\thanks{maxime.lombart@ens-lyon.fr} \\
$^{1}$Univ Lyon, Univ Lyon1, Ens de Lyon, CNRS, Centre de Recherche Astrophysique de Lyon UMR5574, F-69230, Saint-Genis,-Laval, France.\\
$^{2}$Institut Universitaire de France\\
}
\pagerange{\pageref{firstpage}--\pageref{lastpage}} \pubyear{2021}
\date{}
\begin{document}
\label{firstpage}
\bibliographystyle{mnras}
\maketitle

\begin{abstract}
Evolving the size distribution of solid aggregates challenges simulations of young stellar objects. Among other difficulties, generic formulae for stability conditions of explicit solvers provide severe constrains when integrating the coagulation equation for astrophysical objects. Recent numerical experiments have recently reported that these generic conditions may be much too stringent. By analysing the coagulation equation in the Laplace space, we explain why this is indeed the case and provide a novel stability condition which avoids time over-sampling.
\end{abstract}

\begin{keywords}
methods: numerical --- (ISM:) dust, extinction %
\end{keywords}

\section{Introduction}
\label{sec:introduction}

The coagulation equation -- also called Smoluchowski equation -- is one of the fundamental equation of physics, since it describes mass conservation for a distribution of interacting particles \citep{Banasiak2019}. It plays a central role in the formation of planets, since solid bodies, originating from the interstellar medium, have to grow over thirty orders of magnitude in mass to form cores of planets \citep{Chiang2010}. As they grow, grains undergo a complex interplay between coagulation and dynamics since dust/gas interaction depends strongly on the size of the dust grains \citep{Testi2014}. Hence the necessity of performing three dimensional simulations of young stellar objects that integrate the coagulation equation in a self-consistent manner \citep{Haworth2016}. However, this task was long thought to be computationally prohibitive, since no hydrodynamical code could handle the large number of dust bins required to solve for the coagulation equation without over-diffusion. Recently, \citet{Lombart2021} showed that over-diffusivity at small bin numbers could be overcome by the mean of a Discontinuous Galerkine algorithm of high spatial order \citep{Liu2019}. Still, to maintain practical performance, the coagulation solver should not be called too often per hydrodynamical time step \citep{Dra2014,Dra2019}. Stability condition for explicit schemes have been studied since the very beginning of the numerical study of the coagulation equation \citep{FL2004,Dullemond2005,Gabriel2010,Forestier2012,Liu2019}. Time stepping is set by the so-called Courant-Friedrichs-Lewy condition (or CFL condition, \citealt{CFL1928}), which is thought to become drastically small for planet formation, making the solver of no practical use. An alternate solution consists of using implicit solvers, an approach successfully used for fragmentation \citep{Brauer2008,Birnstiel2010,Mahoney2002,Jacobson2005,Sandu2006}, but at the cost of heavy linear algebra operations that increase with the order of the scheme. 

Remarkably, \citet{Liu2019} noticed that his numerical solver was numerically stable for a time step that is orders of magnitude larger than the one given by the generic CFL condition (Sect.~\ref{eq:Smol})  'Only for $\Delta t \leq  0.005$ do we observe a stable solution without using a reconstruction step. This is a significant restriction. With the use of the scaling limiter presented above, we observe that no negative values are generated by the scheme and therefore the solution remains stable, even when raising the time step to $\Delta t = 1.$'. The real CFL condition should therefore be less draconian than the one generically used. Finding it is the goal of this study.

The generic CFL criterion for hyperbolic equations has been proven not only to ensure stability, but also strict positivity of the mass distribution \citep{FL2004}. In a Discontinous Galerkine solver, positivity is instead enforced with a slope limiter, associated with an SSP integrator. Unlocking this positivity constrain ensures numerical stability is actually set by the shortest physical time on which mass transfers through the dust distribution. A high-order scheme ensures then accuracy even when integration is performed with large time steps. Since the coagulation flux is expressed as a double integral over the mass distribution, this time should result from integral considerations over the mass distribution, an information encoded in the Laplace transform of the Smoluchowski equation. Looking at the physical timescales that appear when decomposing the mass distribution into decaying exponentials reveals an alternate and less stringent CFL condition than the ones previously used (Sect.~\ref{eq: CFL}). We validate these findings by testing this condition back in the mass space with the solver of \citet{Lombart2021}. In this study, we focus on constant and additive kernels, since they can be associated with analytic solutions that are the most relevant for planet formation.

\section{Smoluchowski equation}
\label{eq:Smol}

\begin{figure}
\centering
\includegraphics[width=\columnwidth]{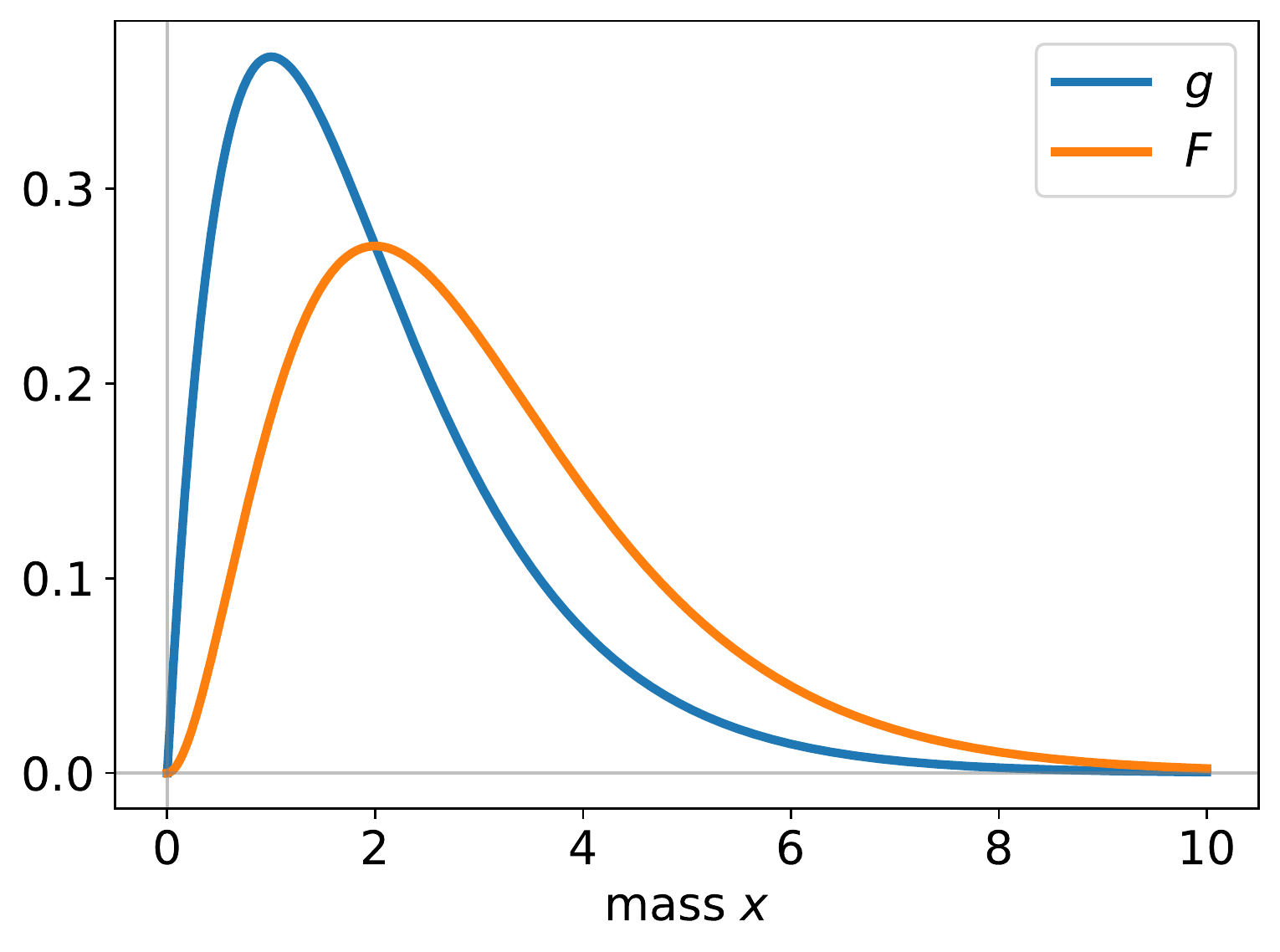}
\caption{Analytic solution for the mass density distribution $g$ and the coagulation flux $F$ for $K = 1$ and $f\left(x,0 \right) = \mathrm{e}^{-x}$ at $t = 1$. The maximum of $g$ does not correspond to the maximum of $F$.}
\label{fig:flux_and_g}
\end{figure}

The Smoluchowski equation is a mean-field, non-linear, integro-differential equation that model mass conservation along a binary collisional process \citep{Smolu1916}. The evolution of the number density of particles per unit mass $f$ is given by
\begin{align}
\frac{\partial f}{\partial t}   = & \frac{1}{2} \int_{0}^{x} K\left(y,x - y \right) f\left( y \right) f\left( x-  y \right) \mathrm{d} y  \nonumber \\
& - f\left( x \right) \int_{0}^{\infty} K\left(y,x \right) f\left( y\right) \mathrm{d}y ,
\label{eq:smolf}
\end{align}
where the kernel $K\left(x,y \right)$ is a symmetric function that gives the collisional rate between particles of mass $x$ and $y$. The conservative form of Eq.~\ref{eq:smolf} is
\begin{equation}
\frac{\partial g}{\partial t} + \frac{\partial F\left[ g \right]}{\partial x} = 0 ,
\label{eq:smolg}
\end{equation}
where $g \equiv xf$ is the mass density distribution per unit mass, and
\begin{equation}
F\left(x \right) = \int_{0}^{x}\int_{x - u}^{\infty} K\left(u,v \right)g\left(u \right)\frac{g\left(v \right)}{v} \mathrm{d}u \mathrm{d}v ,
\label{eq:flux}
\end{equation}
is the coagulation flux \citep{Tanaka1996}. The usual CFL condition for conservative equations of the form Eq.~\ref{eq:smolg} is
\begin{equation}
\frac{\Delta t}{\Delta x}  \max_{g} \left| \frac{\partial F}{\partial g}  \right|  \lesssim 1 ,
\label{eq:CFL_usual}
\end{equation}
For the Smoluchowski equation, the condition given by Eq.~\ref{eq:CFL_usual} may be stringent when considering local individual contribution to the flux of each mass bin. Fig.~\ref{fig:flux_and_g} shows indeed that the quantity $\left| \frac{\partial F}{\partial g}  \right|^{-1}$ can become extremely small, since small increments $\delta g$ may become very small at the location of the maximum of $g$, while $\delta F$ remains finite.  A physical stability condition should instead consider the cumulated contributions of every bins to the local flux, accounting for the contribution of the mass distribution that generates the flux in the mass space. A natural tool to handle these effects consists of determining a stability condition for the time step in the dual Laplace space. We therefore introduce the Laplace transform $\hat{f}\left(p, t \right) \equiv \int_{0}^{\infty} \mathrm{e}^{-p x} f\left(x,t \right) \mathrm{d}x$ of the number density distribution.

\section{Physical time stepping}
\label{eq: CFL}

\subsection{Constant kernel}

\begin{figure}
\centering
\includegraphics[width=\columnwidth]{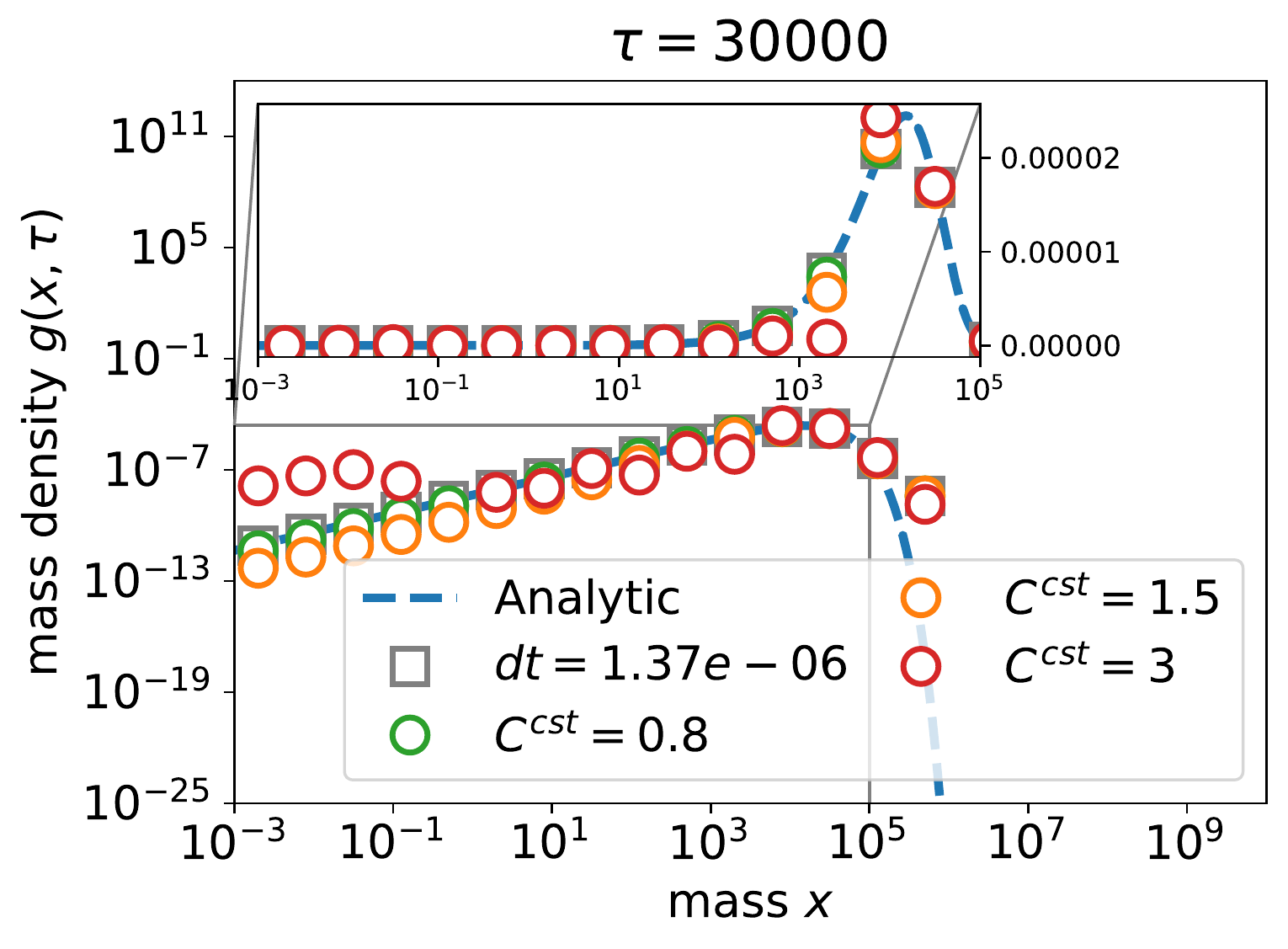}
\caption{Numerical integration of the Smoluchowski equation with a discontinuous Galerkine scheme of order 2. Solutions are displayed in log-log scale for the main plot and lin-log scale when zooming-in close to the maximum. Grey squares correspond to the generic stability condition Eq.~\ref{eq:CFL_usual} (initially, $\Delta t \simeq 1.10^{-6} $). Green, orange and red dotes  correspond to $\Delta t \simeq 0.8 \Delta t _{\rm CFL}$, $\Delta t \simeq 1.5 \Delta t _{\rm CFL}$, $\Delta t \simeq 3 \Delta t _{\rm CFL}$, where $\Delta t _{\rm CFL} = M_{0}^{-1}$ from Eq.~\ref{eq:CFL_kconst_full}. No oscillations develop when the novel CFL condition is satisfied (green), numerical integration remain stable. Blue dashed line: analytic solution.}
\label{fig:compare_kconst}
\end{figure}
We first consider the constant kernel $K = 1$. Taking the Laplace transform of Eq.~\ref{eq:smolf} gives
\begin{equation}
\partial_{t} \hat{f} + M_{0}\left( t \right) \hat{f} - \frac{1}{2}\hat{f}^{2} = 0 ,
\label{eq:Laplace_cst}
\end{equation}
where, for the unit kernel, $M_{0}\left( t \right) = \left( 1 + t/2\right)^{-1}$ \citep{Muller1928}. We first note that
\begin{equation}
0 \le \hat{f}\left( p, t\right) \le \hat{f}\left( 0, t \right)  = M_{0}\left( t \right) .
\label{eq:ineg_fhat}
\end{equation}
Eq.~\ref{eq:ineg_fhat} shows that the non-linear contribution lightens the linear term in Eq.~\ref{eq:Laplace_cst}. Discretising Eq.~\ref{eq:Laplace_cst} with a forward Euler scheme and performing a linear stability analysis of the form $\hat{f}^{n} = \hat{f}_{0}^{n} + \epsilon^{n}$ gives
\begin{equation}
\frac{\epsilon^{n + 1} - \epsilon^{n}}{\Delta t} + M_{0}^{n} \epsilon^{n} = \hat{f}_{0}^{n} \epsilon^{n}, 
\end{equation}
i.e.
\begin{equation}
\epsilon^{n + 1} =   \epsilon^{n} \left[ 1 - \Delta t \left( M_{0}^{n} - \hat{f}_{0}^{n} \right) \right]  .
\end{equation}
Stability of the scheme is obtained at any time from the sufficient condition $\Delta t \le \min \left( M_{0}^{n} -  \hat{f}_{0}^{n}\right)^{-1}$. A stringent majorant for $\Delta t $ is therefore
\begin{equation}
\Delta t \le 1 / M_{0}^{n}.
\label{eq:CFL_add}
\end{equation}
A stability condition for an explicit scheme is therefore set by the typical timescale $M_{0}^{-1}$ of the linear term, as one would expect from the evolution of the moments of the equation \citep{Banasiak2019}. Generally, the novel stability condition should be weighted by a safety coefficient $C^{\rm cst}$
\begin{equation}
\Delta t \le \frac{C^{\rm cst} }{ M_{0}^{n} } =  \mathcal{O} \left( M_{0}^{-1} \right).
\label{eq:CFL_kconst_full}
\end{equation}
This basal Von Neumann analysis is checked by integrating Eq.~\ref{eq:Laplace_cst} within the Laplace space with a forward Euler scheme (Fig.~\ref{fig:laplace_cst}). Numerical results are compared to an analytic solution of Eq.~\ref{eq:Laplace_cst} (integrating only with respect to time in this particuliar case)
\begin{equation}
\hat{f}\left(p,t \right) = \frac{2}{2 + t} ,
\end{equation}
and $C^{\rm cst} = 1$. Numerical results are in excellent agreement with the theoretical analysis. The real test consists of testing the condition Eq.~\ref{eq:CFL_kconst_full} in the mass space. We therefore solve the Smoluchowski equation with the algorithm of \citet{Lombart2021}. We use $9$ orders of magnitude in mass and  $n = 15$ log-spaced bins to mimic the challenging integration conditions encountered in practice. We find numerical stability for the same exact condition (Fig.~\ref{fig:compare_kconst}). For $\Delta t \le \Delta t_{\rm CFL}$, the numerical integration follows the analytic solution with an unexpected accuracy even close to marginal stability, confirming the observation of \citealt{Liu2019}. When $\Delta t \ge \Delta t_{\rm CFL}$, numerical solution breaks strongly at small masses. In practice, the constant $C^{\rm cst}$ should be chosen to provide the desired trade-off between computational efficiency and numerical accuracy. We verified the criterion on distributions with other values of $M_{0}$. 

\subsection{Additive kernel}

\begin{figure}
\centering
\includegraphics[width=\columnwidth]{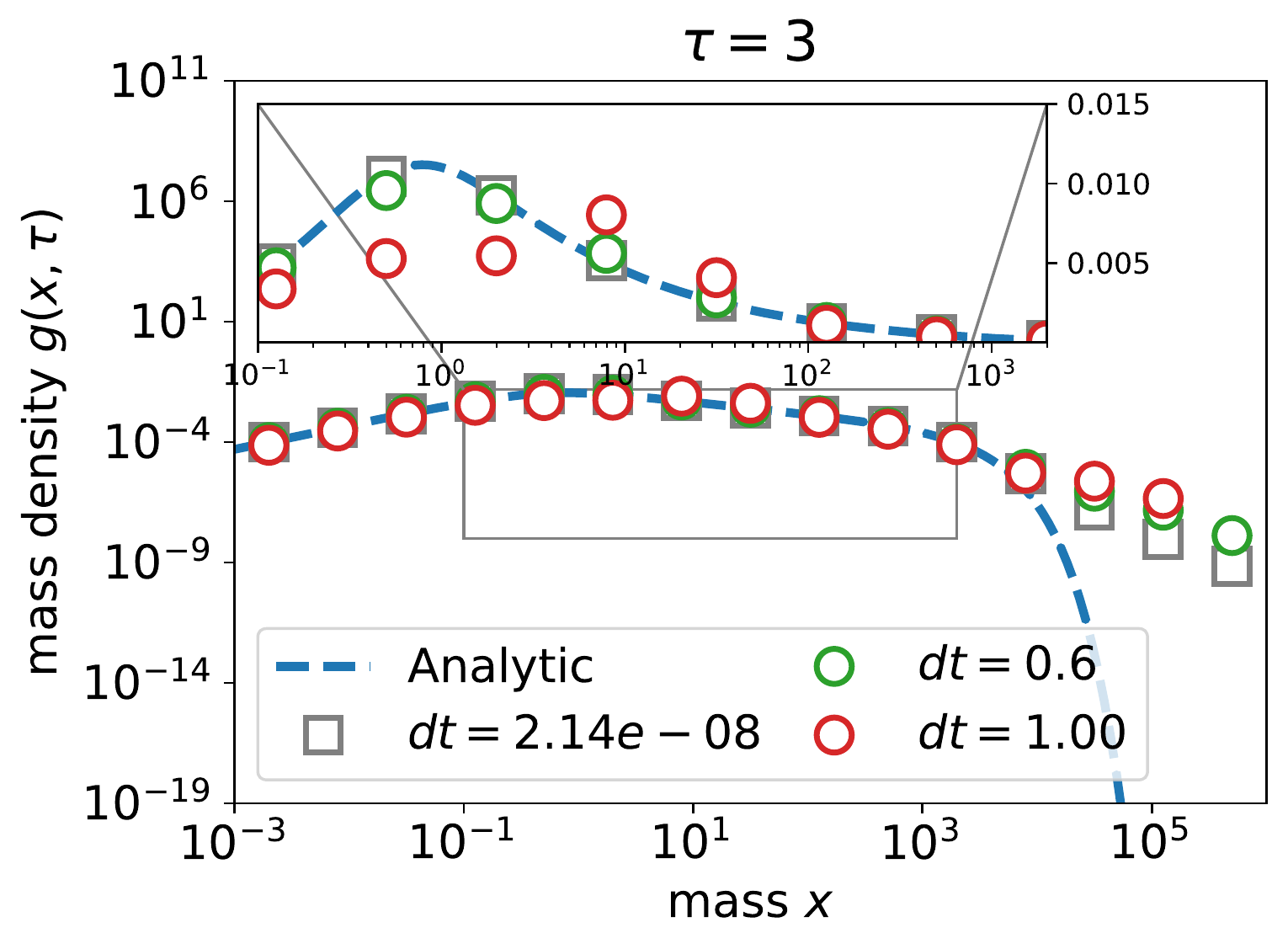}
\caption{Numerical integration of the Smoluchowski equation with a discontinuous Galerkine scheme of order 2. Solutions are displayed in log-log scale for the main plot and lin-log scale when zooming-in close to the maximum. Grey squares correspond to the generic stability condition Eq.~\ref{eq:CFL_usual} (initially, $\Delta t \simeq 2.10^{-8} $). Green dotes and red dots correspond to $\Delta t \simeq 0.9 \Delta t _{\rm CFL}$, $\Delta t \simeq 1.5 \Delta t _{\rm CFL}$, where $\Delta t _{\rm CFL} = 2/3$ from Eq.~\ref{eq:CFL_add3}. No oscillations develop when the novel CFL condition is satisfied (green), numerical integration remain stable. Blue dashed line: analytic solution.}
\label{fig:Laplace_add}
\end{figure}

Let us now consider the additive kernel $K = x + y$. For this kernel, $M_{1}$ is constant and $\frac{\mathrm{d}M_{0}}{\mathrm{d}t} = -M_{1} M_{0}$ \citep{Golovin1963}, hence
\begin{equation}
M_{0}\left( t \right) = M_{0}^{\rm i} \mathrm{e}^{- M_{1} t} .
\label{eq:M1_add}
\end{equation}
The Laplace transform of Eq.~\ref{eq:smolf} is
\begin{equation}
\partial_{t} \hat{f} = \left[ M_{0}\left(t  \right) - \hat{f} \right] \partial_{p} \hat{f} - M_{1}  \hat{f} .
\label{eq:Laplace_add}
\end{equation}
Contrary to the constant case, the contribution of the term $\left[ M_{0}\left(t  \right) - \hat{f} \right] \partial_{p} \hat{f}$ reinforces the contribution of the term $ - M_{1}  \hat{f}$ and contributes to numerical stability. The strategy of analysis consists now in looking at the characteristics of the problem, to show the existence of a real number $C > 0$ that does not depend on $p$, such that $-C \le \partial_{t} \ln \hat{f} \le 0$. As such, $\hat{f}$ decreases slower that a decaying exponential for which the stability condition is known. Eq.~\ref{eq:Laplace_add} becomes
\begin{equation}
\left( \partial_{t} \hat{f} , \partial_{p} \hat{f} , -1 \right) \cdot \left(  1 ,  \hat{f} - M_{0}\left(t  \right) , -M_{1}  \hat{f}   \right) = 0 .
\label{eq:caract_form}
\end{equation}
Eq.~\ref{eq:caract_form} is solved by a method of characteristics by setting $t = t\left( r,s \right)$, $p = p\left( r,s \right)$, $u = u\left( r,s \right) \equiv \hat{f}\left( t, p\right)$, $u\left( t\left( 0,s \right), p\left( 0,s \right) \right) = \hat{f}_{0} \left( s \right)$, following \citet{Banasiak2019}. One has
\begin{align}
 \partial_{r} t & =  1 ,&& t\left(0,s  \right) = 0 ,  \label{eq:c1}\\
 \partial_{r} p & = u - M_{0}\left( r \right) , && p\left(0,s  \right) = s ,  \label{eq:c2}\\
 \partial_{r} u & =  -M_{1} u , &&  u\left(0,s  \right) = \hat{f}_{0} \left( s \right) . \label{eq:c3}
\end{align}
Eq.~\ref{eq:c1} gives $t\left( r,s \right) = r$, Eq.~\ref{eq:c3} gives $u\left( r,s \right) =  \hat{f}_{0} \left( s \right) \mathrm{e}^{-M_{1}r} $. Integrating Eq.~\ref{eq:M1_add}, solving for Eq.~\ref{eq:c2} gives
\begin{equation}
p\left( r, s\right) = s + \frac{\left( \hat{f}_{0} \left( s \right) - M_{0}^{\rm i} \right)}{M_{1}} \left( 1 - \mathrm{e}^{-M_{1} r} \right) .
\label{eq:impl_z}
\end{equation}
Consider now $z\left(r, p \right)$ the implicit solution of Eq.~\ref{eq:impl_z} where $r$ and $p$ are seen as two independent variables, i.e.
\begin{equation}
p = z\left(r, p \right) + \frac{\left( \hat{f}_{0} \left( z\left(r, p \right) \right) - M_{0}^{\rm i} \right)}{M_{1}} \left( 1 - \mathrm{e}^{-M_{1} r} \right) .
\end{equation}
Then, $\hat{f}\left(t\left( r,s \right) , p\left( r,s \right)  \right) = u\left( t, s = z\left(t,p \right) \right)$, and $ \hat{f}\left( t, p\right)$ is expressed on the implicit form $\hat{f}\left( t, p\right) = \hat{f}_{0} \left(z\left(t,p \right)  \right) \mathrm{e}^{-M_{1} t } $. Deriving with respect to time gives
\begin{equation}
\partial_{t} \ln \hat{f} = -M_{1} + \partial_{t} z\left(t,p \right) \frac{\hat{f}'_{0} \left(z\left(t,p \right)  \right)}{\hat{f}_{0} \left(z\left(t,p \right)  \right)} ,
\end{equation}
where we have denoted for convenience $\hat{f}'_{0} \left( p \right) = \partial_{p} \hat{f}_{0} \left( p \right) $. Differentiating Eq.~\ref{eq:impl_z} with respect to $r$ gives
\begin{equation}
\frac{\partial z}{\partial r} = M_{1} \frac{\left(M_{0}^{i} -  \hat{f}_{0}  \right)\mathrm{e}^{-M_{1}r}}{M_{1} +  \hat{f}'_{0} \left( 1 - \mathrm{e}^{-M_{1}r}\right)} .
\end{equation}
The identity $ \hat{f}'_{0}\left( p \right) = - \int_{0}^{\infty} x \mathrm{e}^{- p x} f\left( x \right) \mathrm{d}x $ ensures that
\begin{equation}
0 \le \partial_{t} z \le M_{1} \frac{M_{0}^{i} -  \hat{f}_{0}}{M_{1} +  \hat{f}'_{0} } ,
\end{equation}
and that for any $p$,
\begin{equation}
0 \ge \partial_{t} \ln \hat{f} \ge -M_{1} \left( 1 +  T \left( p\right)  \right).
\end{equation}
where $T \left( p\right) \ge 0$ is given by
\begin{equation}
T \left( p\right)  \equiv \frac{\int_{0}^{\infty} x \mathrm{e}^{- p x} f_{0}\left( x \right) \mathrm{d}x}{\int_{0}^{\infty}  \mathrm{e}^{- p x} f_{0}\left( x \right) \mathrm{d}x}  \frac{\int_{0}^{\infty} \left( 1 - \mathrm{e}^{- p x} \right) f_{0}\left( x \right) \mathrm{d}x}{\int_{0}^{\infty}   \left( 1 - \mathrm{e}^{- p x} \right) x  f_{0}\left( x \right) \mathrm{d}x} .
\end{equation}
Therefore the physical solution decays more slowly than an enveloppe with exponential decay and is associated with the stability condition
\begin{equation}
\Delta t \lesssim \frac{C^{\rm add}}{M_{1}  \left( 1 + \displaystyle \sup_{p} T\left[  f_{0} \right]  \right)   } .
\label{eq:CFL_add}
\end{equation}
In Appendix~\ref{app:bounding}, we prove that $T \left( p\right) \le 1$, allowing us to write the condition of Eq.~\ref{eq:CFL_add}
\begin{equation}
\Delta t \lesssim \frac{C^{\rm add}}{2 M_{1}    } = \mathcal{O} \left( M_{1}^{-1} \right) .
\label{eq:CFL_add2}
\end{equation}
A refined criterion can be obtained when $T\left[  \hat{f}_{0}  \right]$ is actually a decreasing function of $p$. In this case, $T\left( p \right) \le T\left( 0 \right) = \frac{M_{1}^{2}}{M_{0}^{\rm i} M_{2}^{\rm i}} $, which would provide the refined stability condition 
\begin{equation}
\Delta t \lesssim \frac{C^{\rm add}}{M_{1}\left( 1 + \frac{M_{1}^{2}}{M_{0}^{\rm i} M_{2}^{\rm i}}  \right)    }  .
\label{eq:CFL_add3}
\end{equation}
We obtain excellent agreement for the conditions given by Eqs.~\ref{eq:CFL_add} -- \ref{eq:CFL_add3} in the Laplace space (Fig.~\ref{fig:laplace_cst}), against a numerical solution obtained at high resolution. Fig.~\ref{fig:Laplace_add} shows very good applicability of this condition in the real space (varying $M_{1}$ gives similar results). This validates the findings of \citet{Liu2019}. On this example, a factor $\sim 10$ in processing time is gained with the novel condition.\\

The term $\sup_{p} T\left[  f_{0} \right] $ of Eq.~\ref{eq:CFL_add} is the mathematical consequence of the fact that the contribution of $\left[ M_{0}\left(t  \right) - \hat{f} \right] \partial_{p} \hat{f}$ reinforces the the one of $ - M_{1}  \hat{f}$ in Eq.~\ref{eq:Laplace_add}. Finding this correction to be of order unity is physically consistent with fluxes of mass of similar intensities generated by the two terms of the right-hand-side of Eq.~\ref{eq:smolf}. We note that the CFL condition comes from the limit $p\to 0$, which corresponds to the limit case of a constant mass distribution that is non-integrable over the mass space. Mass fluxes are indeed expected to be more intense for this distribution, since an additive kernel favour growth over the largest grains. We conjecture that this CFL condition can, alternatively, be obtained from the evolution of the moments of Eq.~\ref{eq:smolf}. The method presented here can be applied to other relevant coagulation kernels.

\section{Conclusion}
\label{sec:conclusion}

In this study, we revisit the derivation of the stability condition for explicit numerical solvers of the Smoluchowski equation. Generic formulae are too stringent since they also ensure positivity. Enforcing positivity by some alternate way (e.g. with a slope limiter associated to an SSP integrator) allows to improve stability condition by several orders of magnitude. Novel conditions that involve moments of the mass distribution -- $\Delta t_{\rm CFL} \sim M_{0}^{-1}$ for the constant kernel, $\Delta t_{\rm CFL} \sim M_{1}^{-1}$ for the additive kernel -- are obtained by analysing dual problems in the Laplace space, to account for the non-locality of the coagulation equation and its different responses to different dust distributions. Numerical simulations are in excellent agreement with the theory and validate our novel CFL condition, confirming the observations of \citet{Liu2019}. 

\section*{Acknowledgements}

GL acknowledges funding from the ERC CoG project PODCAST No 864965. This project has received funding from the European Union's Horizon 2020 research and innovation programme under the Marie Sk\l odowska-Curie grant agreement No 823823. This project was partly supported by the IDEXLyon project (contract nANR-16-IDEX-0005) under the auspices University of Lyon. We acknowledge financial support from the national programs (PNP, PNPS, PCMI) of CNRS/INSU, CEA, and CNES, France. We thank E. Lynch for useful comments.

\section*{Data availability}
\label{data_github}
The data and supplementary material underlying this article are available in the repository "growth" on GitHub at \url{https://github.com/mlombart/growth.git}. Figures can be reproduced following the file \texttt{README.md}. The repository contains data and Python scripts used to generate figures.

\label{lastpage}

\bibliography{biblio,biblio_num_smo}

\begin{appendix}
\onecolumn
\section{Bounding of $T$}
\label{app:bounding}

Following \citet{Yang2017}, let denote
\begin{equation}
T \left( p\right)  \equiv \underbrace{\frac{\int_{0}^{\infty} x \mathrm{e}^{- p x} f_{0}\left( x \right) \mathrm{d}x}{\int_{0}^{\infty}  \mathrm{e}^{- p x} f_{0}\left( x \right) \mathrm{d}x}}_{T_{1} \equiv I_{1} / I_{2}} \underbrace{ \frac{\int_{0}^{\infty} \left( 1 - \mathrm{e}^{- p x} \right) f_{0}\left( x \right) \mathrm{d}x}{\int_{0}^{\infty}   \left( 1 - \mathrm{e}^{- p x} \right) x  f_{0}\left( x \right) \mathrm{d}x}}_{T_{2} \equiv I_{3} / I_{4}} .
\end{equation}
Deriving $T_{1}$ with respect to $p$ and symmetrising $x \leftrightarrow y $ gives
\begin{align}
I_{2}\left(p \right)^{2} \frac{\mathrm{d} T_{1}\left[  f \right] \left( p \right)}{\mathrm{d} p} & \equiv - \left\lbrace  \left( \int_{0}^{\infty} \mathrm{e}^{- p x} f\left( x \right) \mathrm{d}x \right)  \left( \int_{0}^{\infty} \mathrm{e}^{- p x} x^{2}f\left( x \right) \mathrm{d}x \right) -  \left( \int_{0}^{\infty} \mathrm{e}^{- p x} x f\left( x \right) \mathrm{d}x \right)^{2}  \right\rbrace , \\
& = - \left \lbrace   \int_{0}^{\infty} \!\!  \int_{0}^{\infty} \mathrm{d}x \mathrm{d}y \, \mathrm{e}^{-px} \mathrm{e}^{-py} f\left(x \right) \! f\left(y \right) y^{2}   -  \int_{0}^{\infty} \!\!  \int_{0}^{\infty} \mathrm{d}x \mathrm{d}y \, \mathrm{e}^{-px} \mathrm{e}^{-py} f\left(x \right) \! f\left(y \right) xy    ,    \right \rbrace\\
& = - \left \lbrace   \int_{0}^{\infty} \!\!  \int_{0}^{\infty} \mathrm{d}x \mathrm{d}y \, \mathrm{e}^{-px} \mathrm{e}^{-py} f\left(x \right) \! f\left(y \right) \frac{\left( x^{2} + y^{2} \right)}{2}   -  \int_{0}^{\infty} \!\!  \int_{0}^{\infty} \mathrm{d}x \mathrm{d}y \, \mathrm{e}^{-px} \mathrm{e}^{-py} f\left(x \right) \! f\left(y \right) x y      \right \rbrace  ,\\
& = - \frac{1}{2} \int_{0}^{\infty} \!\!  \int_{0}^{\infty} \mathrm{d}x \mathrm{d}y \, \mathrm{e}^{-px} \mathrm{e}^{-py} f\left(x \right) \! f\left(y \right) \left( x - y \right)^{2} < 0 .
\label{eq:deriv_T1}
\end{align}
Hence, $T_{1}\left[  f \right]$ is a decreasing function for any $f$. As such,
\begin{equation}
T_{1}\left[  f_{0}  \right]  \left( p \right) \le T_{1}\left[  f_{0}  \right]  \left( 0 \right) = \frac{M_{1}}{M_{0}} .
\label{eq:sup_T1}
\end{equation}
Similarly, deriving $T_{2}$ with respect to $p$ gives
\begin{align}
I_{4}\left(p \right)^{2} \frac{\mathrm{d} T_{2}\left[  f \right] \left( p \right)}{\mathrm{d} p} & \equiv  \left\lbrace  \left( \int_{0}^{\infty}  \left( 1 - \mathrm{e}^{- p x} \right) x f\left( x \right) \mathrm{d}x \right)  \left( \int_{0}^{\infty} x \mathrm{e}^{- p x} f\left( x \right) \mathrm{d}x \right) \right. \nonumber \\ 
& \left. \,\,\,\,\,\,\,\,\,\,\,\,\,\,\,\, -  \left( \int_{0}^{\infty}  x^2 \mathrm{e}^{- p x} f\left( x \right) \mathrm{d}x \right)  \left( \int_{0}^{\infty} \left( 1 - \mathrm{e}^{- p x} \right) f\left( x \right) \mathrm{d}x \right)  \right\rbrace \\
& =  \left\lbrace  \int_{0}^{\infty} \!\!  \int_{0}^{\infty} \mathrm{d}x \mathrm{d}y \, f\left(x \right) \! f\left(y \right) x y \left(1 - \mathrm{e}^{-px} \right)\mathrm{e}^{-py}       -  \int_{0}^{\infty} \!\!  \int_{0}^{\infty} \mathrm{d}x \mathrm{d}y \, f\left(x \right) \! f\left(y \right) x^2 \mathrm{e}^{-px} \left(1 - \mathrm{e}^{-py} \right) \right\rbrace ,\\
& = \frac{1}{2} \left\lbrace  \int_{0}^{\infty} \!\!  \int_{0}^{\infty} \mathrm{d}x \mathrm{d}y \, f\left(x \right) \! f\left(y \right) \underbrace{ \left( y - x\right) \left[ x \mathrm{e}^{-px} \left(1 - \mathrm{e}^{-py} \right) - y \mathrm{e}^{-py}\left(1 - \mathrm{e}^{-px} \right) \right] }_{\ge 0 } \right\rbrace .
\label{eq:deriv_T2}
\end{align}
Hence, $T_{2}\left[  f \right]$ is strictly increasing function for any $f$,
\begin{equation}
T_{2}\left[  f_{0}  \right]  \left( p \right) \le T_{2}\left[ f_{0}  \right]  \left( \infty \right) = \frac{M_{0}}{M_{1}} .
\label{eq:sup_T2}
\end{equation}
Finally, $T \left( p\right)  = T_{1}  \left( p\right) T_{2}  \left( p\right)\le 1$. 

\section{Stability condition in the Laplace space}
\label{app:Laplace}
\begin{figure}
\centering
\includegraphics[width=0.45\columnwidth]{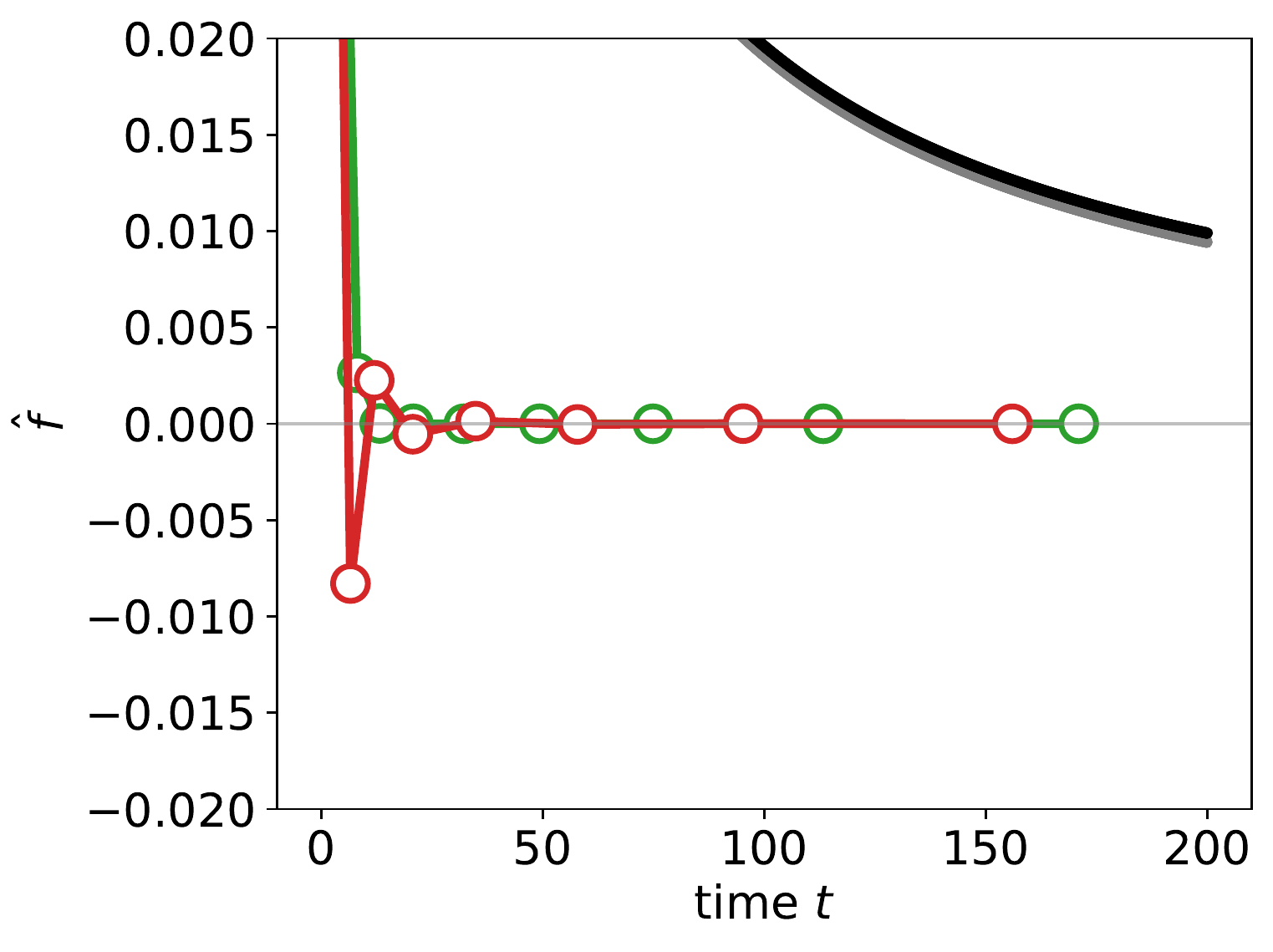}
\includegraphics[width=0.45\columnwidth]{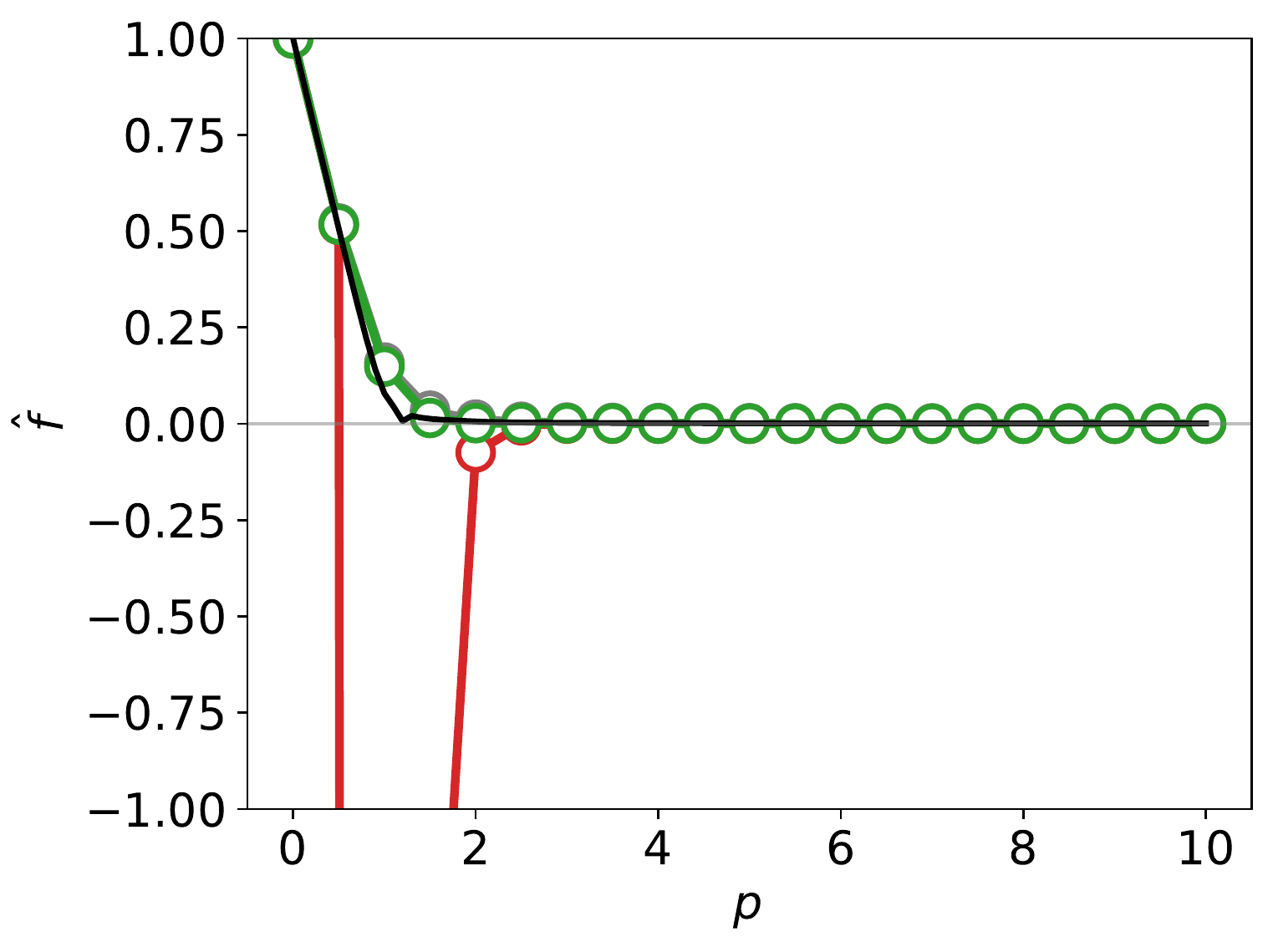}
\caption{Left: Numerical solution of Eq.~\ref{eq:Laplace_cst} obtained with a first order Euler scheme. Small grey, large green and large red dots correspond to $\Delta t = 10^{-3} \Delta t _{\rm CFL}$, $\Delta t = 1. \Delta t _{\rm CFL}$ and $\Delta t = 1.25 \Delta t _{\rm CFL}$, where $\Delta t _{\rm CFL} $ is given by Eq.~\ref{eq:CFL_add} ($C^{\rm cst} = 1$). $\Delta t$ varies with time. Black solid line: analytic solution. Right: Numerical solution of Eq.~\ref{eq:Laplace_add} obtained with a first order upwind scheme, under the condition $\hat{f}\left( 0 ,t \right) = 1$ with 20 grid points. Grey, large green and large red dots correspond to $\Delta t \simeq 0.09 \Delta t _{\rm CFL}$, $\Delta t \simeq 0.83 \Delta t _{\rm CFL}$ and $\Delta t \simeq 1.07 \Delta t _{\rm CFL}$, where $\Delta t _{\rm CFL} = 2/3$ from Eq.~\ref{eq:CFL_add3} ($C^{\rm add} = 1$). Black solid line: analytic solution (a better sampling in mass makes the numerical solution closer to the analytic solution).}
\label{fig:laplace_cst}
\end{figure}
Fig.~\ref{fig:laplace_cst} show the validity of the stability conditions obtained for the numerical integration of $\hat{f}$ (Laplace space) in Sect.~\ref{eq: CFL} .

\end{appendix}

\end{document}